\documentclass[twocolumn,english,aps,prl,floatfix,amssymb,superscriptaddress,longbibliography]{revtex4}
\usepackage[latin9]{inputenc}
\usepackage{verbatim}
\usepackage{float}
\usepackage{amsmath}
\usepackage{amssymb}
\usepackage{graphicx}
\usepackage{color}
\usepackage{xcolor}
\usepackage{tikz}
\newcommand{\ket}[1]{\ensuremath{\left| #1 \right>}}

\usepackage{bbold}

\usepackage[bookmarks=false,linkcolor=blue,urlcolor=blue,colorlinks,citecolor=blue]{hyperref}
\makeatletter

\newcommand{\be}{\begin{equation}}
\newcommand{\ee}{\end{equation}}
\newcommand{\bea}{\begin{eqnarray}}
\newcommand{\eea}{\end{eqnarray}}

\begin{document}

\title{Quantum Integrability of Hamiltonians with Time-Dependent Interaction Strengths and the Renormalization Group Flow}
\author{Parameshwar R. Pasnoori}
\affiliation{Department of Physics, University of Maryland, College Park, MD 20742, United
States of America}
\email{pparmesh@umd.edu}
\begin{abstract}
In this paper we consider quantum Hamiltonians with time-dependent interaction strengths, and following the recently formulated generalized Bethe ansatz framework [P. R. Pasnoori, Phys. Rev. B 112, L060409 (2025)], we show that constraints imposed by integrability take the same form as the renormalization group flow equations corresponding to the respective Hamiltonians with constant interaction strengths. As a concrete example, we consider the anisotropic time-dependent Kondo model characterized by the time-dependent interaction strengths $J_{\parallel}(t)$ and $J_{\perp}(t)$. We construct an exact solution to the time-dependent Schrodinger equation and by applying appropriate boundary conditions on the fermion fields we obtain a set of matrix difference equations called the quantum Knizhnik-Zamolodchikov (qKZ) equations corresponding to the XXZ R-matrix. The consistency of these equations imposes constraints on the time-dependent interaction strengths $J_{\parallel}(t)$ and $J_{\perp}(t)$, such that the system is integrable. Remarkably, the resulting temporal trajectories of the couplings are shown to coincide exactly with the RG flow trajectories of the static Kondo model, establishing a direct and universal correspondence between integrability and renormalization-group flow in time-dependent quantum systems.

\end{abstract}
\maketitle

Renormalization group (RG) theory offers a unifying framework for understanding how the behavior of a physical system changes as one moves between energy or length scales \cite{Stueckelberg,gellmanlow}. By systematically integrating out short-distance fluctuations \cite{Wilson1,Wilson}, one arrives at RG flow equations which govern the scale dependence of couplings encoded in the $\beta$-functions \cite{callan,Symanzik}. These flows classify interactions by their relevance, identify fixed points that exhibit scale invariance, and explain the universality shared by microscopically distinct models \cite{kadanoff1,kogut}. In quantum integrable systems, however, the RG structure is tightly constrained: the preservation of an infinite tower of conserved quantities often forces $\beta$-functions to vanish when the system exhibits conformal invariance, or it confines trajectories to narrow manifolds \cite{Dolan} compatible with factorized scattering and absence of particle production \cite{ZAMOLODCHIKOV,Zamolodchikov1,Dorey,FATEEV1987,BanksZaks1982,Novikov1983,Polchinski1988}. In the case of quantum integrable Hamiltonians with constant interaction strengths, Bethe ansatz framework provides the exact solution including the eigenstate spectrum and also the exact RG flow equations of the coupling strengths \cite{Bethe1931,SklyaninQISM,McGuire,GAUDIN,AndreiLowensein81,LL1,LL2,Thacker}.

For quantum integrable models with time-dependent interaction strengths, the generalized Bethe ansatz framework \cite{PasnooriKondo,PasnooriGrossNeveu} uniquely fixes the admissible functional forms of the interaction strengths compatible with integrability, and also provides the exact time-dependent wavefunction that is the solution to the time-dependent Schrodinger equation. In this work we establish that these integrability conditions are exactly equivalent to the renormalization-group flow equations of the corresponding Hamiltonian with time-independent interactions. This result reveals a direct and nontrivial identification between the two descriptions, obtained by mapping the physical time $t$ of the driven system onto the logarithmic cutoff scale of the static problem $t=\log \Lambda$, thereby uncovering a fundamental link between integrability and RG flow in Hamiltonians with time-dependent coupling strengths. This connection was first established in the context of integrable sigma models with coupling strengths that are dependent on the world-sheet time \cite{Benhoare}.  It was shown that the integrability is preserved, that is the time-dependent theory admits a Lax connection provided the time-dependent coupling strengths satisfy the one-loop RG equations of the time-independent theory.

%For time-dependent interaction strengths, integrability uniquely fixes the admissible functional forms, permitting only those temporal dependences compatible with integrability \cite{PasnooriKondo}. We establish that these integrability conditions are exactly equivalent to the renormalization-group flow equations of the corresponding Hamiltonian with time-independent interactions. This result reveals a direct and nontrivial identification between the two descriptions, obtained by mapping the physical time $t$ of the driven system onto the logarithmic cutoff scale of the static problem $t=\log \Lambda$, thereby uncovering a fundamental link between integrability and RG flow in Hamiltonians with time-dependent coupling strengths.

In contrast to the sigma models, integrability of the quantum field theories such as the $SU(N)$ Gross-Neveu model and quantum impurity models such as the Kondo model is established by constructing the exact Bethe ansatz wavefunction which provides the exact S-matrices that satisfy the quantum Yang-Baxter equations. We illustrate that these class of Bethe ansatz integrable systems also exhibit the connection between the time-dependent coupling strengths and the RG flows of the respective time-independent models by considering a paradigmatic quantum impurity model with two couplings strengths, namely the anisotropic Kondo model.  The Hamiltonian of the anisotropic Kondo model is given by
\begin{align}\nonumber
H\hspace{-0.5mm}&=\hspace{-1.2mm} \int_{0}^L\hspace{-1mm} dx \left(-i\psi^{\dagger}_a(x)\partial_x\psi_a(x)+ \delta(x) \psi^{\dagger}_a(x)\psi_b(x)\; \vec{A}_{ab,\alpha\beta}(t)\right),\\&\vec{A}_{ab,\alpha\beta}(t)=\frac{1}{2}J_{\parallel}(t) \sigma^z_{ab}S^z_{\alpha\beta}+
\frac{1}{2}J_{\perp}(t) (\sigma^x_{ab}S^x_{\alpha\beta}+\sigma^y_{ab}S^y_{\alpha\beta}).\label{kondoham}\end{align}

Here, $\psi_a(x)$ is a fermion field with spin $a=\uparrow\downarrow$ and $S$ represents the impurity. $J_{\parallel}(t)$ and $J_{\perp}(t)$ are time dependent coupling strengths corresponding to the interaction between the impurity and the electrons close to the impurity. The first term aligns the spin of the electrons and that of the impurity along the $z$-direction, while the second term is responsible for exchanging their spins. The system exhibits a $U(1)$ symmetry in the charge sector which corresponds to the conservation the total number of electrons $N$, where
\begin{align} N=\sum_a\int_0^L dx\; \psi^{\dagger}_a(x)\psi_a(x).
\end{align}
The system also exhibits a $U(1)$ symmetry in the spin sector, which corresponds to the conservation of the spin along the $z$-direction, as
$S^z_{\rm T}=  s^z + S^z$ commutes with the Hamiltonian (\ref{kondoham}). Here $S^z $ is the z-component of the impurity spin and $s^z = \int_{0}^L dx\; s^z(x)$ is that of the electrons, where
\be
\label{spinz}
 s^z(x)= \frac{1}{2} (\Psi^{\dagger}_{a}(x)  \sigma^z_{ab} \Psi^{}_{b}(x) ).
\ee
In the special limit $J_{\parallel}(t)=J_{\perp}(t)$, the system exhibits an enlarged $SU(2)$ symmetry and has been solved using Bethe ansatz method \cite{Andrei80,Wiegmann_1981}.  

The case where the interaction strengths $J_{\parallel}(t)$ and $J_{\perp}(t)$ are constants is solvable by the standard Bethe ansatz method \cite{TsvelickWiegmann1983}. For values of $J_{\parallel}\geq J_{\perp}$, the system is renormalizable and results in the generation of a strong coupling energy scale called the Kondo temperature, below which, the impurity forms a many body singlet state with conduction electrons resulting in the screening of the impurity and increase in the resistivity \cite{JunKondo,Wilson,Nozier}. Here we consider the Hamiltonian (\ref{kondoham}) with time dependent interaction strengths and solve it exactly using the generalized Bethe ansatz method formulated in \cite{PasnooriKondo}. In contrast to the conventional Bethe ansatz, which is limited to Hamiltonians with time-independent interactions, the generalized Bethe ansatz provides a unified framework for treating systems with explicitly time-dependent interaction strengths. It enforces integrability constraints on the admissible temporal profiles of the interaction strengths and, when these conditions are met, the framework yields exact solutions of the time-dependent Schrodinger equation. This framework was applied to solve the Hamiltonian (\ref{kondoham}) in the $SU(2)$ limit $J_{\parallel}(t)=J_{\perp}(t)$. It was later used in \cite{PasnooriGrossNeveu} to solve the $SU(2)$ Gross-Neveu model with time-dependent interaction strength.

Below, following \cite{PasnooriKondo}, we construct the exact wavefunction and derive the constraints on the interaction strengths imposed by integrability. As the number of particles is a conserved quantity, one can construct the wavefunction that is labeled by $N$. The time-dependent Schrodinger equation takes the form 
\begin{align} H\ket{\Psi}_N=i\partial_t\ket{\Psi}_N,\label{SEk}
\end{align}
where $H$ is the Hamiltonian (\ref{kondoham}). 

\paragraph{One particle sector}
Let us first consider the case where we have one particle $N=1$. We have
\begin{align}
\ket{\Psi}_1=\int_0^L dx \; \psi^{\dagger}_a(x)F_{a\alpha}(x,t)\ket{\alpha},\end{align}
were $\alpha$ represents the spin of the impurity. Using this in the Schrodinger equation (\ref{SEk}) one obtains the following one particle Schrodinger equation 
\begin{align}-i(\partial_t+\partial_x)F_{a\alpha}(x,t)+\delta(x) \vec A_{ab,\alpha\beta}(t) F_{b\beta}(x,t)=0.\label{se1p}
\end{align}
The first term describes the motion of the particle with linear dispersion. The second term describes the interaction of the particle with the impurity. Before considering the general solution to the above equation (\ref{se1p}), let us consider the situation in which the impurity is absent. In the absence of the impurity, the above equation (\ref{se1p}) only has the first term
\begin{align}-i(\partial_t+\partial_x)F_{a}(x,t)=0. \;\;\;\text{(No impurity case)}\label{se1pni}
\end{align}
Any function of the following form is a solution
\begin{align}F_{a}(x,t)=f_a(x-t). \;\; \text{(No impurity case)}\label{sol1ni}\end{align} 
Now consider the case where the impurity is present. Since the particle only interacts with the impurity at $x=0$, the particle is free when it is away from the impurity for all $x\neq 0$. Hence, away from the impurity, the wavefunction should be of the form (\ref{sol1ni}). As the particle interacts with the impurity locally, one may look for a wavefunction of the form where the amplitudes corresponding to the particle being on the left and right sides of the impurity is different
\begin{align}
    F_{a\alpha}(x,t)=f^{10}_{a\alpha}(x-t)\theta(-x)+f^{01}_{a\alpha}(x-t)\theta(x).
\end{align}
Here $\theta(x)$ is the Heaviside function with $\theta(x)=1, x>0$ and $\theta(x)=0, x<0$. The superscripts on the amplitudes $f^{10}_{a\alpha}(x-t)$ and $f^{01}_{a\alpha}(x-t)$ denote the ordering of the particle with respect to the impurity. These amplitudes are related to each other through the particle-impurity S-matrix
\begin{align}f^{01}_{a\alpha}(z)=S^{10}_{ab,\alpha\beta}(z)f^{10}_{b\beta}(z),\label{1prel}
\end{align}
where we have used the notation $z=x-t$. Here the summation of the repeated indices is assumed. The explicit form of the S-matrix $S^{10}_{ab,\alpha\beta}(z)$ takes the following form
\begin{align}\label{smatk}
  S^{10}(t)= \begin{pmatrix} 1&0&0&0\\0&\frac{\sinh(h(t))}{\sinh(h(t)+iu(t))}&i\frac{\sin(u(t))}{\sinh(h(t)+iu(t))}&0\\0&i\frac{\sin(u(t))}{\sinh(h(t)+iu(t))}&\frac{\sinh(h(t))}{\sinh(h(t)+iu(t))}&0\\0&0&0&1\end{pmatrix},
\end{align}
where $h(t)$ and $u(t)$ are related to the coupling strengths $J_{\parallel}(t)$ and $J_{\perp}(t)$ through the relations

\begin{align}\nonumber h(-t)&=\text{arccoth}\left\{\left(\frac{\sin{\frac{J_{\parallel}(t)}{2}}/\sin(\frac{1}{2}(J_{\parallel}(t)-J_{\perp}(t)))}{\sin(\frac{1}{2}(J_{\parallel}(t)+J_{\perp}(t))}\right)^{1/2}\right\}\\u(-t)&=\arccos\left(\frac{\cos(J_{\parallel}(t)/2)}{\cos(J_{\perp}(t)/2)}\right).\label{relint}
\end{align}
These cumbersome equations have an interpretation in terms of the RG flow equations, as we shall see later. We have seen that the two amplitudes corresponding to the particle being on the left and right sides of the impurity are related to each other through the particle-impurity S-matrix such that there exists one free amplitude. To determine the exact form of the complete wavefunction, we need to determine this amplitude. This can be achieved by applying periodic boundary conditions on the fermion fields $\psi(x)=\psi(x+L)$. This results in the following relation between the amplitudes
\begin{align}f^{10}_{a\alpha}(z-L)=f^{01}_{a\alpha}(z).\label{1pbck}
\end{align}
Using the above equation (\ref{1pbck}) in (\ref{1prel}), we arrive at the following relation

\bea \label{1pdiffk}f^{10}_{a\alpha}(z-L)=S^{10}_{ab\alpha\beta}(z)f^{10}_{b\beta}(z).\eea
This is a matrix difference equation which can be solved to obtain the explicit form of the amplitude $f^{10}_{a\alpha}(z)$. Using the relation (\ref{1prel}) or (\ref{1pbck}), one can then obtain the other amplitude $f^{01}_{a\alpha}(z)$, and hence the explicit form of the complete wavefunction.

\paragraph{N-particle sector}
Now let us consider the case of $N>1$ particles. The wavefunction takes the form
\begin{align}\label{npform}\ket{\Psi_N}=\hspace{-0.13in}\sum_{\sigma_1...\sigma_N}\hspace{-0.5mm}\prod_{j=1}^{N}\int_{0}^L\hspace{-0.5mm}dx_j \Psi^{\dagger}_{\sigma_j}(x_j)\mathcal{A}F_{\sigma_1...\sigma_N\alpha}(x_1,...,x_N,t)\ket{\alpha}, 
\end{align}
where $\sigma_j$ denote the spin indices of the electrons and $\alpha$ denotes the spin of the impurity and $\mathcal{A}$ denotes anti-symmetrization with respect to $x_i$ and $\sigma_i$. Using the above expression (\ref{npform}) in the Schrodinger equation (\ref{SEk}), one obtains the following equation
\begin{align} &\nonumber-i(\partial_t+\sum_{j=1}^N\partial_{x_j})F_{\sigma_1...\sigma_N\alpha}(x_1,...,x_N,t)+
\\&\big(\sum_{j=1}^N\delta(x_j)\vec A_{\sigma_j\gamma_j,\alpha\beta}(t)\hspace{-1.5mm}\prod_{l=1,l\neq j}^N\hspace{-2mm}I_{\sigma_l\gamma_l}\big)F_{\gamma_1...\gamma_N\beta}(x_1,...,x_N,t)=0.\label{npsek}\end{align}
Here $I_{\sigma_l\gamma_l}$ is the identity matrix. In the $N>1$ particle case, in addition to distinguishing the amplitudes corresponding to the ordering of the particles with respect to the impurity, integrability requires one to distinguish the amplitudes which differ by the ordering of the particles with respect to each other. From here on we suppress the spin indices unless required, for the ease of notation. The $N$-particle wavefunction takes the following form
\begin{align} \nonumber 
F_{\sigma_1...\sigma_N}(x_1,...x_N,t)
= \sum_Q \theta(\{x_{Q(j)}\})  f^Q_{\sigma_1...\sigma_N}(z_1,...z_N).
\end{align}
In this expression, $Q$ denotes a permutation of the position orderings of particles and  $\theta(\{x_{Q(j)}\})$ is the Heaviside function that vanishes unless $x_{Q(1)} < \dots < x_{Q(N)}$. Here $f^{Q}_{\sigma_1...\sigma_N} (z_1,...,z_N)$ is the amplitude corresponding to the ordering of the particles denoted by $Q$. The two amplitudes corresponding to a particle $j$ being on the left and right sides of the impurity are related through the particle-impurity S-matrix (\ref{smatk})

\bea \label{nprel}f^{..0j..}(z)=S^{j0}(z)\:f^{..j0..}(z).\eea
 
 Here ``.." in the superscripts corresponds to any ordering of the rest of the particles, which is the same in both the amplitudes. As mentioned above, the integrability requires one to choose the particle-particle S-matrices which relate the amplitudes $f^{..ij..}(z_1,...,z_N)$ and $f^{..ji..}(z_1,...,z_N)$ that differ by the ordering of the two particles $i$ and $j$ with respect to each other
 \begin{align}
f^{..ji.}(z_1,...,z_N) =S^{ij}(z_i,z_j)f^{..ij..}(z_1,...,z_N).\label{eemateqsn}\end{align}
  Here again ``.." in the superscripts corresponds to any ordering of the rest of the particles, which is the same in both amplitudes. This is required even in the case of constant interaction strength for the consistency of the wave function. These pairs of amplitudes are not constrained by the Hamiltonian due to the relativistic dispersion \cite{PasnooriKondo}. These S-matrices are chosen such that they satisfy the following Yang-Baxter's equation \cite{BAXTER,CNYang} with the particle impurity S-matrices
\begin{align}S^{j0}(z_j)S^{i0}(z_i)S^{ij}(z_i,z_j)=S^{ij}(z_i,z_j)S^{i0}(z_i)S^{j0}(z_j)\label{YB2}.\end{align}
This imposes a restriction on the interaction strengths $J_{\parallel}(t)$ and $J_{\perp}(t)$ such that 
\begin{align}\cos(J_{\parallel}(t&))=\cos(u)\cos(J_{\perp}(t)),\label{constraint1}\\&\text{(First constraint)}\nonumber
\end{align}
where $u(-t)\equiv u$ is a constant. The only consistent form of the particle-particle S-matrix $S^{ij}(z_i,z_j)$ that satisfies the Yang-Baxter relation (\ref{YB2}) takes the following form
\begin{align} S^{ij}(z_i,z_j)=\begin{pmatrix} 1&0&0&0\\0&\frac{\sinh(h(z_i)-h(z_j))}{\sinh(h(z_i)-h(z_j)+iu)}&\frac{i\sin(u)}{\sinh(h(z_i)-h(z_j)+iu)}&0\\0&\frac{i\sin(u)}{\sinh(h(z_i)-h(z_j)+iu}&\frac{\sinh(h(z_i)-h(z_j))}{\sinh(h(z_i)-h(z_j)+iu)}&0\\0&0&0&1\end{pmatrix}\label{smatee}
\end{align}

The particle-particle S-matrices also satisfy the Yang-Baxter equation

\begin{align}\nonumber
&S^{ij}(z_i,z_j)S^{ik}(z_i,z_k)S^{jk}(z_j,z_k)
\\&=S^{jk}(z_j,z_k)S^{ik}(z_i,z_k)S^{ij}(z_i,z_j).\label{YB3}
\end{align}
Applying periodic boundary conditions results in the following relations
\begin{align}
    f^{j...}_{\sigma_1...\sigma_N}(z_1,...,z_j,...,z_N)=f^{...j}_{\sigma_1...\sigma_N}(z_1,...,z_j+L,...,z_N), \label{pbcnp}   \end{align}
where again ``..." in the superscripts corresponds to any ordering of the rest of the particles, which is the same in both amplitudes. Using the relations (\ref{nprel}), (\ref{eemateqsn}) and the Yang-Baxter relations (\ref{YB2}) and (\ref{YB3}), one can express any amplitude in the N-particle wavefunction in terms of one amplitude of our choosing. Let us choose this amplitude to be
$f^{N...10}_{\sigma_1...\sigma_N}(z_1,...,z_N)$. To obtain the explicit form of the N-particle wavefunction, we need to determine this amplitude. Just as in the one-particle case, this can be achieved by applying periodic boundary conditions (\ref{pbcnp}). This results in the following matrix difference equation
\begin{align} \nonumber &f^{N,...,j,...10}_{\sigma_1...\sigma_N}(z_1,..z_j-L,..,z_N)\\&=Z_j(z_1,...,z_N) \; f^{N,...j,...10}_{\sigma_1...\sigma_N}(z_1,..,z_j,..,z_N),\label{qKZ}\end{align}

where

\begin{align}\nonumber
    Z_j(z_1,...,z_N)= S^{jj+1}(z_j,z_{j+1}+L)S^{jj+2}(z_j,z_{j+2}+L)\\...S^{jN}(z_j,z_N+L)S^{j0}(z_j)S^{j1}(z_j,z_1)...S^{jj-1}(z_j,z_{j-1})
.\label{transfermat}\end{align}
The operator $Z_{j}(z_1,...,z_N)$ transports the particle $j$ through the entire system once, when acting on the amplitude $f^{N...10}_{\sigma_1...\sigma_N}(z_1,...,z_N)$. Integrability requires that transporting a particle $`i'$ throughout the system and then transporting the particle $`j'$ should be equivalent to transporting particle $`j'$ first and then transporting the particle $`i'$. This gives rise to the following constraint equation on the transport operator (\ref{transfermat})
\begin{align}\nonumber
&Z_i(z_1,...,z_j-L,...,z_N) Z_j(z_1,...,z_N)\\&=
Z_j(z_1,...,z_i-L,...,z_N) Z_i(z_1,...,z_N).\label{commutations}\end{align}
\paragraph{Integrable interaction strengths}
We have seen that for the wavefunction to be consistent, the interaction strengths $J_{\parallel}(t)$ and $J_{\perp}(t)$ are restricted by the relation (\ref{constraint1}). This nevertheless does not constrain the exact functional form of $J_{\parallel}(t)$, or equivalently $J_{\perp}(t)$. This is precisely fixed by the constraint equations (\ref{commutations}), which results in

\begin{align}
  h(t\pm L)=h(t)\pm\kappa ,\label{constraint2}\\\nonumber\text{(Second constraint)}  \end{align}
where $\kappa$ is a constant and $h(t)$ is related to $J_{\parallel}(t)$ and $J_{\perp}(t)$ through the relation (\ref{relint}). 
The constraint condition (\ref{constraint2}) is satisfied when $h(t)$ takes the following form
\begin{align}h(t)=at+c,\label{formf}
\end{align}
were $a=\kappa/L$ and $c$ is a constant \footnote{$a$ and $c$ can be complex which correspond to non-Hermitian Hamiltonian}. Using (\ref{constraint2}) in (\ref{relint}), we obtain the functional forms of the interaction strengths $J_{\parallel}(t)$ and $J_{\perp}(t)$ which give rise to an integrable system. They take the following form

\begin{align}\nonumber J_{\perp}(t)=2\arccos\left(\frac{\pm\tanh(at+c)}{\left(\sin^2(u)+\cos^2(u)\tanh^2(at+c)\right)^{1/2}}\right)\\J_{\parallel}(t)=2\arccos\left(\frac{\pm\cos(u)\tanh(at+c)}{\left(\sin^2(u)+\cos^2(u)\tanh^2(at+c)\right)^{1/2}}\right).\label{formints}\end{align}
These functional forms of the time-dependent integrable interaction strengths fix the exact form of the particle-impurity and particle-particle S-matrices (\ref{smatk}) and (\ref{smatee}), and hence also the transport operator (\ref{transfermat}). For interaction strengths satisfying the integrability constraints (\ref{constraint1}) and (\ref{constraint2}), the matrix difference equations (\ref{qKZ}) turn into \textit{quantum Knizhnik-Zamolodchikov} (qKZ) equations. The solutions to the qKZ equations have been well studied in the literature \cite{Smirnov_1986,Frenkel,rishetikhin1,rishetikhin2,Babujian_1997,JimboMiwa1995,Cherednik2006,Tarasov:1994bb}, using which one can obtain the explicit form of the exact wavefunction. For our current discussion, the explicit form of the wavefunction is not relevant and hence it will be presented elsewhere. 
\paragraph{Constraints imposed by integrability and their relation to the renormalization group flow}
 
Consider the constraints imposed by integrability on the time-dependent interaction strengths $J_{\parallel}(t)$ and $J_{\perp}(t)$, which is given in (\ref{constraint1}). For small values of $J_{\parallel}(t)$ and $J_{\perp}(t)$, this equation takes the form
\begin{align}
4u^2=J_{\parallel}(t)^2-J_{\perp}(t)^2. \label{rg1}
\end{align}
Recall the relation between $f(t)$ and the interaction strengths $J_{\parallel}(t)$ and $J_{\perp}(t)$ provided in the first line of (\ref{relint}). Using (\ref{rg1}), this can be rewritten as
\begin{align}
   \text{coth}(h(t))=\frac{J_{\parallel}(t)}{2u}. \label{rg2}
\end{align}
 Differentiating both sides of the above equation with respect to `$t$', and using the relation (\ref{rg1}), we obtain
\begin{align}\left(1-\frac{J_{\parallel}(t)^2}{4u^2} \right)\frac{d}{dt} f(t)=\frac{1}{2u}\frac{d}{dt}J_{\parallel}(t).\label{rg3}
\end{align}
Using the explicit form of $h(t)$ (\ref{formf}) in the above equation, we obtain
\begin{align}
\frac{d}{dt}J_{\parallel}(t)=\frac{a}{2u}J_{\perp}(t)^2.\label{rg4}
\end{align}
Where, we have used (\ref{rg1}). Now Differentiating (\ref{rg1}) with respect to `$t$' on both sides and using the above equation, we obtain
\begin{align}
\frac{d}{dt}J_{\perp}(t)=\frac{a}{2u}J_{\parallel}(t)J_{\perp}(t). \label{rg5}
\end{align}
The equations (\ref{rg4}) and (\ref{rg5}) look very similar to renormalization group (RG) equations of the anisotropic Kondo model with constant interaction strength \cite{TsvelickWiegmann1983}. The equations (\ref{rg4}) and (\ref{rg5}) exactly coincide with the RG equations, provided
\begin{align}
   t=\log(\Lambda), \;\; a=-2u/\pi, \label{timeenergy} 
\end{align}
where $\Lambda$ is the cutoff. With the simple identification of time and energy cutoff in (\ref{timeenergy}), we find that the integrability constraints in (\ref{constraint1}) and (\ref{constraint2}) and their seemingly cumbersome solutions for the integrable interaction strengths in (\ref{formints}) exhibit a strikingly elegant structure: Integrability of a Hamiltonian with time-dependent interaction strengths requires that the temporal evolution of the interaction strengths follow trajectories that are identical to the renormalization-group flows of the corresponding Hamiltonian with time-independent interactions. This nontrivial correspondence between RG flow and time-dependent integrability persists across different regularization schemes, underscoring its robustness. To illustrate this explicitly, we focus on the $SU(2)$ limit.

\paragraph{The $SU(2)$ limit}

 The $SU(2)$ limit corresponds to $J_{\parallel}(t)=J_{\perp}(t)=J(t)$. In this limit $u\rightarrow 0$ (\ref{relint}), where the particle-impurity and particle-particle S-matrices take the form of the XXX-R matrix \cite{PasnooriKondo}, where the parameter $f(t)$ is related to the interaction strength $J(t)$ through the relation
\begin{align} h(-t)=\cot( J(t)/2).
\end{align}

We emphasize that no specific regularization of the Heaviside function $\theta(x)$ has been assumed. Adopting a particular choice, such as $\theta(0)=1/2$, modifies the relation between $h(t)$ and the interaction strength $J(t)$ \cite{PasnooriKondo}. Nevertheless, the integrability condition remains universally fixed by (\ref{constraint2}) in all regularization schemes. Consequently, upon inverting the relation between $h(t)$ and $J(t)$, the explicit functional form of the integrable interaction strength $J(t)$ depends on the chosen regularization, while the underlying integrability constraint itself is unchanged.

 An analogous subtlety arises for time-independent interactions, $J(t)=J$.  In the Bethe ansatz approach, the Kondo temperature depends on the parameter $h(t)=h$ according to 
 \begin{align}T_K=\Lambda e^{-\pi h} \;\; \text{Bethe ansatz}.\label{Bethetk}
\end{align}
Since the relation between $h$ and the interaction strength $J$ is regularization dependent, expressing 
$T_K$ in terms of $J$ yields different functional forms for different regularization schemes. At first sight, this appears paradoxical, as physical observables must be independent of the regularization procedure. Moreover, consistency demands that the Kondo temperature assume the same functional dependence as that obtained from the Wilson renormalization-group analysis,
\begin{align}T_K=\Lambda e^{-\pi/J}\;\; \text{Wilson (one loop)}.\label{Wilsontk}
\end{align}

These apparent ambiguities are resolved by restricting attention to the universal weak-coupling regime, in which physical observables become independent of the regularization scheme. For the $SU(2)$ Kondo model, this regime corresponds to small coupling strengths $J$, where both regularization schemes give rise to the same relation between the parameter $f$ and the interaction strength $J$, where $h=1/J$. In this regime, the resulting Kondo temperature $T_K$ assumes the universal form in (\ref{Bethetk}), in exact agreement with the one-loop Wilson renormalization-group result in (\ref{Wilsontk}).

The same reasoning extends to the case of time-dependent interaction strengths. In this setting, the applicability of different regularization schemes is restricted to regimes in which they yield the same universal functional form for the time-dependent coupling. This corresponds to asymptotic time scales for which the relation between the parameter $f(t)$ and the interaction strength
$J(t)$ becomes unique and independent of regularization. In the long-time limit, all regularization schemes converge to the same universal time-dependent interaction strength
\begin{align}J(t)=\frac{\pi}{t}.\label{formintsu2}
\end{align}
 Note that the equations (\ref{rg4}) and  (\ref{rg5}) take the same form in the $SU(2)$ limit
\begin{align}
\frac{d}{dt}J(t)=-\frac{1}{\pi}J(t)^2.\label{rg6}
\end{align}

As anticipated, upon invoking the identification in (\ref{timeenergy}), this equation reduces to the renormalization-group flow of the Kondo model in the $SU(2)$ limit \cite{TsvelickWiegmann1983}. The resulting functional form of the interaction strength in (\ref{formintsu2}) satisfies (\ref{rg6}), thereby establishing that imposing regularization-scheme independence uniquely fixes the coupling strength's functional form, with its time evolution governed by the renormalization-group equation.
Finally, note that from (\ref{formints}) one can see that in the Toulouse limit \cite{Toulouse1969}, which corresponds to $u=\pi/2$, only $J_{\perp}(t)$ flows, whereas $J_{\parallel}(t)=0$ for all times. This is in agreement with the respective RG flows \cite{TsvelickWiegmann1983}.

\paragraph{Discussion}
For integrable Hamiltonians with time-independent interactions, the constraints imposed by integrability ensure that the form of the Hamiltonian is preserved across energy scales, with the coupling strengths undergoing renormalization governed by the renormalization-group equations. Considering the anisotropic Kondo model, which is a paradigmatic quantum impurity system with multiple couplings, we have demonstrated that, when the interactions are made time dependent, integrability imposes stringent constraints on their allowed functional forms. Specifically, the Hamiltonian remains integrable at all times provided the interaction strengths evolve along temporal trajectories that exactly coincide with the corresponding renormalization-group flow trajectories of the corresponding Hamiltonian with time-independent interaction strengths.

We conjecture that this correspondence between renormalization-group flow and time evolution of interaction strengths is generic. More broadly, for any integrable Hamiltonian with constant couplings, integrability is preserved under time-dependent driving if and only if the couplings follow their respective renormalization-group equations, with the logarithmic cutoff scale $\log\Lambda$ identified with the physical time $t$.

\section*{Acknowledgements}
I acknowledge discussions with P. Azaria and A. Tsvelik. I thank P. Azaria for carefully reviewing the manuscript.

\bibliography{refpaper}
\end{document}